\documentstyle[preprint,floats,pra,eqsecnum,aps,psfig]{revtex}
\tightenlines
\begin{document} \draft

\title{Covariant Model of Relativistic Extended Particles based
on the Oscillator Representation of the Poincar\'e Group}

\author{Y. S. Kim}
\address{Department of Physics, University of Maryland, College Park,
Maryland 20742}

\maketitle

\begin{abstract}

While internal space-time symmetries of relativistic particles are
dictated by the little groups of the Poincar\'e group, it is possible
to construct representations of the little group for massive particles
starting from harmonic oscillator wave functions for the quark model.
The resulting oscillator wave functions are covariant and can be
Lorentz-boosted.  It is thus possible to obtain the parton model by
boosting the quark model.  A review of Wigner's theory of the little
groups is given.  It is shown that the covariant oscillator wave
functions become squeezed as as the system becomes boosted.  It is
shown also that the Lorentz-squeezed quark distribution exhibits the
peculiarities of Feynman's parton model including the lack of coherence
in the calculation of cross sections.  A historical review of the
concept of covariance is given.

\end{abstract}


\section{Introduction}\label{intro}
The present form of quantum mechanics works well in atomic systems
where electrons are bound by the Coulomb force from the nucleus.
Quantum mechanics works also for atomic nuclei where the nucleus is
a bound state of nucleons even though it is difficult to perform exact
calculations.  In both atomic and nuclear physics, nucleons are regarded
as point particles.  However, it was found by Hofstadter in
1995~\cite{hofsta55} that the proton's charge is not concentrated on a
point, and its
charge distribution has a non-zero space-time extension.  This appears
in the elastic scattering of electrons by a proton.  The observed
scattering cross sections deviate from those predicted by the
Rutherford scattering formula based on the proton with a point charge.
This deviation is commonly called the form factor.

In spite of many laudable efforts to explain this form factor within
the framework of quantum field theory, the workable model for the form
factors did not emerge until after Gell-Mann's formulation of the quark
model, in which all hadrons are bound states of quarks and/or
anti-quarks~\cite{gell64}.  The question then is whether we can use
the existing models of quantum mechanics, such as the nuclear shell
model, to explain hadronic mass spectra~\cite{owg64}.  For the mass
spectra, one of the most effective models has been and still is the
model based on harmonic oscillator wave functions~\cite{owg64,fkr71}.
The basic advantage of the oscillator model is that its mathematics is
transparent, and it does not bury physics in mathematics even though
it does not always produce the most accurate numerical results.

In the quark model, the charge distribution within the proton comes
from the distribution of the charged particles inside the hadron.
The success of the oscillator model for static or slow-moving
hadrons does not necessarily mean that the model can be extended to
the relativistic regime.  Indeed, the calculation of the form factor
with Gaussian wave functions results in an exponential decrease for
large momentum-transfer variables.  However, this wrong behavior
comes from the use of non-relativistic wave functions for relativistic
problems.  Indeed, Feynman {\it et al.} made an attempt to construct a
covariant oscillator model~\cite{fkr71}.  Even though they did not
achieve their goal in their paper, Feynman {\it et al.} quote the work
of Fujimura {\it et al.}~\cite{fuji70} who calculated the nucleon form
factor by taking into account the effect of the Lorentz-squeeze on the
oscillator wave functions.

After studying these original papers, we can raise our level of
abstraction.  We observe first that the spherical harmonics can represent
the three-dimensional rotation group, while serving as wave functions for
the angular variables.  Then, we can ask whether there are wave functions
which can represent the Poincar\'e group.  We can specifically ask whether
it is possible to construct a set of normalizable harmonic oscillator
wave functions to represent the Poincar\'e group.  If YES, the wave
functions can be Lorentz-boosted.  These wave functions then have to go
through another set of tests.  Are they consistent with the existing laws
of quantum mechanics.  If YES, they then have to be exposed to the most
cruel test in physics.  Do they explain what we observe in high-energy
laboratories?

The purpose of this paper is to show that we can use the oscillator
wave functions to answer the question of whether quarks are partons.
While the quark model is valid for static hadrons, Feynman's parton
picture works only in the Lorentz frame where the hadronic speed is
close to that of light~\cite{fey69}.
The quark model appears to be quite different from the parton model.
On the other hand, they are valid in two different Lorentz frames.
The basic question is whether the quark picture and the parton picture
are two different manifestations of the same covariant entity.

In this paper, we shall discuss first the internal space-time symmetries
of relativistic particles in terms of appropriate representations of
the Poincar\'e group~\cite{wig39}.  We then construct the oscillator
wave functions satisfying the above-mentioned theoretical criterions.
This oscillator formalism will explains both the quark and the parton
pictures in two separate Lorentz frames.  This formalism produces all
the peculiarities of Feynman's original form of the parton picture
including the incoherence of parton cross sections.

In Sec.~\ref{littleg}, we present a brief history of applications of
the little groups to internal space-time symmetries of relativistic
particles.  In Sec.~\ref{covham}, we construct representations of
the little group using harmonic oscillator wave functions.
In Sec.~\ref{parton}, it is shown that the Lorentz-boosted oscillator
wave functions exhibit the peculiarities Feynman's parton model in
the infinite-momentum limit.

Much of the concept of Lorentz-squeezed wave function is derived from
elliptic deformations of a sphere resulting in a mathematical technique
group called contractions~\cite{inonu53}.  In Appendix~\ref{o3e2}, we
discuss the contraction of the three-dimensional rotation group to the
two-dimensional Euclidean group.  In Appendix~\ref{contrac}, we
discuss the little group for a massless particle as the
infinite-momentum/zero-mass limit of the little group for a massive
particle.  In Appendix~\ref{kant}, the author gives his confession
about his educational and cultural backgrounds which led to the
research program outlined in this paper.

\section{Little Groups of the Poincar\'e Group}\label{littleg}

The Poincar\'e group is the group of inhomogeneous Lorentz
transformations, namely Lorentz transformations preceded or followed
by space-time translations.  In order to study this group, we have to
understand first the group of Lorentz transformations, the group of
translations, and how these two groups are combined to form the
Poincar\'e group.  The Poincar\'e group is a semi-direct product of
the Lorentz and translation groups.  The two Casimir operators of
this group correspond to the (mass)$^{2}$ and (spin)$^{2}$ of a given
particle.  Indeed, the particle mass and its spin magnitude are
Lorentz-invariant quantities.

The question then is how to
construct the representations of the Lorentz group which are relevant to
physics.  For this purpose, Wigner in 1939 studied the subgroups of the
Lorentz group whose transformations leave the four-momentum of a given free
particle \cite{wig39}.  The maximal subgroup of the Lorentz group
which leaves the four-momentum invariant is called the little group.
Since the little group leaves the four-momentum invariant, it governs the
internal space-time symmetries of relativistic particles.  Wigner shows in
his paper that the internal space-time symmetries of massive and massless
particles are dictated by the $O(3)$-like and $E(2)$-like little groups
respectively.

The $O(3)$-like little group is locally isomorphic to the three-dimensional
rotation group, which is very familiar to us.  For instance, the group
$SU(2)$ for the electron spin is an $O(3)$-like little group.  The group
$E(2)$ is the Euclidean group in a two-dimensional space, consisting
of translations and rotations on a flat surface.  We are performing
these transformations everyday on ourselves when we move from home to
school.  The mathematics of these Euclidean transformations are also
simple.  However, the group of these transformations are not well
known to us.  In Appendix \ref{o3e2}, we give a matrix representation
of the $E(2)$ group.

The group of Lorentz transformations consists of three boosts and
three rotations.  The rotations therefore constitute a subgroup of
the Lorentz group.  If a massive particle is at rest, its four-momentum
is invariant under rotations.  Thus the little group for a massive
particle at rest is the three-dimensional rotation group.  Then what is
affected by the rotation?  The answer to this question is very simple.
The particle in general has its spin.  The spin orientation is going
to be affected by the rotation!

If the rest-particle is boosted along the $z$ direction, it will pick
up a non-zero momentum component.  The generators of the $O(3)$ group
will then be boosted.  The boost will take the form of conjugation by
the boost operator.  This boost will not change the Lie algebra of the
rotation group, and the boosted little group will still leave the
boosted four-momentum invariant.  We call this the $O(3)$-like little
group.  If we use the four-vector coordinate $(x, y, z, t)$, the
four-momentum vector for the particle at rest is $(0, 0, 0, m)$, and
the three-dimensional rotation group leaves this four-momentum invariant.
This little group is generated by
\begin{equation}
J_{1} = \pmatrix{0&0&0&0\cr0&0&-i&0\cr0&i&0&0\cr0&0&0&0} , \qquad
J_{2} = \pmatrix{0&0&i&0\cr0&0&0&0\cr-i&0&0&0\cr0&0&0&0} ,
\end{equation}
and
\begin{equation}\label{j3}
J_{3} = \pmatrix{0 & -i & 0 & 0 \cr i & 0 & 0 & 0
\cr 0 & 0 & 0 & 0 \cr 0 & 0 & 0 & 0} ,
\end{equation}
which satisfy the commutation relations:
\begin{equation}
[J_{i}, J_{j}] = i\epsilon_{ijk} J_{k} .
\end{equation}

It is not possible to bring a massless particle to its rest frame.
In his 1939 paper~\cite{wig39}, Wigner observed that the little group
for a massless particle moving along the $z$ axis is generated by the
rotation generator around the $z$ axis, namely $J_{3}$ of Eq.(\ref{j3}),
and two other generators which take the form
\begin{equation}\label{n1n2}
N_{1} = \pmatrix{0 & 0 & -i & i \cr 0 & 0 & 0 & 0
\cr i & 0 & 0 & 0 \cr i & 0 & 0 & 0} ,  \quad
N_{2} = \pmatrix{0 & 0 & 0 & 0 \cr 0 & 0 & -i & i
\cr 0 & i & 0 & 0 \cr 0 & i & 0 & 0} .
\end{equation}
If we use $K_{i}$ for the boost generator along the i-th axis, these
matrices can be written as
\begin{equation}
N_{1} = K_{1} - J_{2} , \qquad N_{2} = K_{2} + J_{1} ,
\end{equation}
with
\begin{equation}
K_{1} = \pmatrix{0&0&0&i\cr0&0&0&0\cr0&0&0&0\cr i&0&0&0} , \qquad
K_{2} = \pmatrix{0&0&0&0\cr0&0&0&i\cr0&0&0&0\cr0&i&0&0} .
\end{equation}
The generators $J_{3}, N_{1}$ and $N_{2}$ satisfy the following set
of commutation relations.
\begin{equation}\label{e2lcom}
[N_{1}, N_{2}] = 0 , \quad [J_{3}, N_{1}] = iN_{2} ,
\quad [J_{3}, N_{2}] = -iN_{1} .
\end{equation}
In Appendix \ref{o3e2}, we discuss the generators of the $E(2)$ group.
They are $J_{3}$ which generates rotations around the $z$ axis, and
$P_{1}$ and $P_{2}$ which generate translations along the $x$ and $y$
directions respectively.  If we replace $N_{1}$ and $N_{2}$ by $P_{1}$
and $P_{2}$, the above set of commutation relations becomes the set
given for the $E(2)$ group given in Eq.(\ref{e2com}).  This is the
reason why we say the little group for massless particles is
$E(2)$-like.  Very clearly, the matrices $N_{1}$ and $N_{2}$ generate
Lorentz transformations.

It is not difficult to associate the rotation generator $J_{3}$ with
the helicity degree of freedom of the massless particle.   Then what
physical variable is associated with the $N_{1}$ and $N_{2}$ generators?
Indeed, Wigner was the one who discovered the existence of these
generators, but did not give any physical interpretation to these
translation-like generators.  For this reason, for many years, only
those representations with the zero-eigenvalues of the $N$ operators
were thought to be physically meaningful representations~\cite{wein64}.
It was not until 1971 when Janner and Janssen reported that the
transformations generated by these operators are gauge
transformations~\cite{janner71,kim97poz}.  The role of this
translation-like transformation has also been studied for spin-1/2
particles, and it was concluded that the polarization of neutrinos
is due to gauge invariance~\cite{hks82,kim97min}.

\begin{table}

\caption{Further contents of Einstein's $E = mc^{2}$.  Massive and
massless particles have different energy-momentum relations.  Einstein's
special relativity gives one relation for both.  Wigner's little group
unifies the internal space-time symmetries for massive and massless
particles which are locally isomorphic to $O(3)$ and $E(2)$ respectively.
It is a great challenge for us to find another unification.  In this
note, we present a unified picture of the quark and parton models which
are applicable to slow and ultra-fast hadrons respectively.}

\vspace{3mm}

\begin{tabular}{cccc}

{}&{}&{}&{}\\
{} & Massive, Slow \hspace*{1mm} & COVARIANCE \hspace*{1mm}&
Massless, Fast \\[4mm]\hline
{}&{}&{}&{}\\
Energy- & {}  & Einstein's & {} \\
Momentum & $E = p^{2}/2m$ & $ E = [p^{2} + m^{2}]^{1/2}$ & $E = cp$
\\[4mm]\hline
{}&{}&{}&{}\\
Internal & $S_{3}$ & {}  &  $S_{3}$ \\[-1mm]
space-time &{} & Wigner's  & {} \\ [-1mm]
symmetry & $S_{1}, S_{2}$ & Little Group & Gauge
Transformations \\[4mm]\hline
{}&{}&{}&{}\\
Relativistic & {} & {} & {} \\[-1mm]
Extended & Quark Model & Covariant Model of Hadrons & Partons \\ [-1mm]
Particles & {} & {} & {} \\[2mm]
\end{tabular}
\end{table}

Another important development along this line of research is the
application of group contractions to the unifications of the two
different little groups for massive and massless particles.
We always associate the three-dimensional rotation group with a spherical
surface.  Let us consider a circular area of radius 1 kilometer centered
on the north pole of the earth.  Since the radius of the earth is more
than 6,450 times longer, the circular region appears flat.  Thus, within
this region, we use the $E(2)$ symmetry group for this region.  The
validity of this approximation depends on the ratio of the two radii.

In 1953, Inonu and Wigner formulated this problem as the contraction of
$O(3)$ to $E(2)$~\cite{inonu53}.  How about then the little groups which
are isomorphic to $O(3)$ and $E(2)$?  It is reasonable to expect that the
$E(2)$-like little group be obtained as a limiting case for of the
$O(3)$-like little group for massless particles.  In 1981, it was
observed by Ferrara and Savoy that this limiting process is the Lorentz
boost \cite{ferrara82}.  In 1983, using the
same limiting process as that of Ferrara and Savoy, Han {\it et al}
showed that transverse rotation generators become the generators of
gauge transformations in the limit of infinite momentum and/or zero mass
\cite{hks83pl}.  In 1987, Kim and Wigner showed that the little group for
massless particles is the cylindrical group which is isomorphic to the
$E(2)$ group~\cite{kiwi87jm}.  This completes the second raw in Table I,
where Wigner's little group unifies the internal space-time symmetries
of massive and massless particles.

We are now interested in constructing the third row in Table I.  As we
promised in Sec.~\ref{intro}, we will be dealing with hadrons which are
bound states of quarks with space-time extensions.  For this purpose, we
need a set of covariant wave functions consistent with the existing laws
of quantum mechanics, including of course the uncertainty principle and
probability interpretation.

With these wave functions, we propose to solve the following problem in
high-energy physics.  The quark model works well when hadrons are at
rest or move slowly.  However, when they move with speed close to that
of light, they appear as a collection of infinite-number of
partons~\cite{fey69}.  As we stated above, we need a set of wave
functions which can be Lorentz-boosted.  How can we then construct such
a set?  In constructing wave functions for any purpose in quantum mechanics,
the standard procedure is to try first harmonic oscillator wave functions.
In studying the Lorentz boost, the standard language is the Lorentz group.
Thus the first step to construct covariant wave functions is to work
out representations of the Lorentz group using harmonic
oscillators~\cite{dir45,yuka53,knp86}.

\section{Covariant Harmonic Oscillators}\label{covham}

If we construct a representation of the Lorentz group using normalizable
harmonic oscillator wave functions, the result is the covariant harmonic
oscillator formalism~\cite{knp86}.  The formalism constitutes a
representation of Wigner's $O(3)$-like little group for a massive
particle with internal space-time structure.  This oscillator formalism
has been shown to be effective in explaining the basic phenomenological
features of relativistic extended hadrons observed in high-energy
laboratories.  In particular, the formalism shows that the quark model
and Feynman's parton picture are two different manifestations of one
covariant entity~\cite{knp86,kim89}.  The essential feature of the
covariant harmonic oscillator formalism is that Lorentz boosts are
squeeze transformations~\cite{kn73,knp91}.  In the light-cone coordinate
system, the boost transformation expands one coordinate while contracting
the other so as to preserve the product of these two coordinate remains
constant.  We shall show that the parton picture emerges from this
squeeze effect.

Let us consider a bound state of two particles.  For convenience, we
shall call the bound state the hadron, and call its constituents quarks.
Then there is a Bohr-like radius measuring the space-like separation
between the quarks.  There is also a time-like separation between the
quarks, and this variable becomes mixed with the longitudinal spatial
separation as the hadron moves with a relativistic speed.  There are
no quantum excitations along the time-like direction.  On the other
hand, there is the time-energy uncertainty relation which allows
quantum transitions.  It is possible to accommodate these aspect within
the framework of the present form of quantum mechanics.  The uncertainty
relation between the time and energy variables is the c-number
relation~\cite{dir27},
which does not allow excitations along the time-like coordinate.  We
shall see that the covariant harmonic oscillator formalism accommodates
this narrow window in the present form of quantum mechanics.

For a hadron consisting of two quarks, we can consider their space-time
positions $x_{a}$ and $x_{b}$, and use the variables
\begin{equation}
X = (x_{a} + x_{b})/2 , \qquad x = (x_{a} - x_{b})/2\sqrt{2} .
\end{equation}
The four-vector $X$ specifies where the hadron is located in space and
time, while the variable $x$ measures the space-time separation between
the quarks.  In the convention of Feynman {\it et al.} \cite{fkr71},
the internal motion of the quarks bound by a harmonic oscillator
potential of unit strength can be described by the Lorentz-invariant
equation
\begin{equation}\label{osceq}
{1\over 2}\left\{x^{2}_{\mu} -
{\partial ^{2} \over \partial x_{\mu }^{2}}
\right\} \psi (x)= \lambda \psi (x) .
\end{equation}
It is now possible to construct a representation of the Poincar\'e group
from the solutions of the above differential equation~\cite{knp86}.

The coordinate $X$ is associated with the overall hadronic
four-momentum, and the space-time separation variable $x$ dictates
the internal space-time symmetry or the $O(3)$-like little group.  Thus,
we should construct the representation of the little group from the
solutions of the differential equation in Eq.(\ref{osceq}).  If the
hadron is at rest, we can separate the $t$ variable from the equation.
For this variable we can assign the ground-state wave function to
accommodate the c-number time-energy uncertainty relation~\cite{dir27}.
For the three space-like variables, we can solve the oscillator
equation in the spherical coordinate system with usual orbital and
radial excitations.  This will indeed constitute a representation of
the $O(3)$-like little group for each value of the mass.  The solution
should take the form
\begin{equation}
\psi (x,y,z,t) = \psi (x,y,z) \left({1\over \pi }\right)^{1/4}
\exp \left(-t^{2}/2 \right) ,
\end{equation}
where $\psi(x,y,z)$ is the wave function for the three-dimensional
oscillator with appropriate angular momentum quantum numbers.  Indeed,
the above wave function constitutes a representation of Wigner's
$O(3)$-like little group for a massive particle \cite{knp86}.

Since the three-dimensional oscillator differential equation is
separable in both spherical and Cartesian coordinate systems,
$\psi(x,y,z)$ consists of Hermite polynomials of $x, y$, and $z$.
If the Lorentz boost is made along the $z$ direction, the $x$ and $y$
coordinates are not affected, and can be temporarily dropped from the wave
function.  The wave function of interest can be written as
\begin{equation}
\psi^{n}(z,t) = \pmatrix{{1\over \pi }}^{1/4}\exp \pmatrix{-t^{2}/2}
\psi_{n}(z) ,
\end{equation}
with
\begin{equation}
\psi ^{n}(z) = \left({1 \over \pi n!2^{n}} \right)^{1/2} H_{n}(z)
\exp (-z^{2}/2) ,
\end{equation}
where $\psi ^{n}(z)$ is for the $n$-th excited oscillator state.
The full wave function $\psi ^{n}(z,t)$ is
\begin{equation}\label{2.6}
\psi ^{n}_{0}(z,t) = \left({1\over \pi n! 2^{n}}\right)^{1/2} H_{n}(z)
\exp \left\{-{1\over 2}\left(z^{2} + t^{2} \right) \right\} .
\end{equation}
The subscript $0$ means that the wave function is for the hadron at rest.
The above expression is not Lorentz-invariant, and its localization
undergoes a Lorentz squeeze as the hadron moves along the $z$
direction~\cite{knp86}.

It is convenient to use the light-cone variables to describe Lorentz
boosts.  The light-cone coordinate variables are
\begin{equation}
u = (z + t)/\sqrt{2} , \qquad v = (z - t)/\sqrt{2} .
\end{equation}
In terms of these variables, the Lorentz boost along the $z$
direction,
\begin{equation}
\pmatrix{z' \cr t'} = \pmatrix{\cosh \eta & \sinh \eta \cr
\sinh \eta & \cosh \eta } \pmatrix{z \cr t} ,
\end{equation}
takes the simple form
\begin{equation}\label{lorensq}
u' = e^{\eta } u , \qquad v' = e^{-\eta } v ,
\end{equation}
where $\eta $ is the boost parameter and is $\tanh ^{-1}(v/c)$.
Indeed, the $u$ variable becomes expanded while the $v$ variable becomes
contracted.  This is the squeeze mechanism illustrated discussed
extensively in the literature~\cite{kn73,knp91}.  This squeeze
transformation is also illustrated in Fig.~1.

The wave function of Eq.(\ref{2.6}) can be written as
\begin{equation}\label{10}
\psi ^{n}_{o}(z,t) = \psi ^{n}_{0}(z,t)
= \left({1 \over \pi n!2^{n}} \right)^{1/2} H_{n}\left((u + v)/\sqrt{2}
\right) \exp \left\{-{1\over 2} (u^{2} + v^{2}) \right\} .
\end{equation}
If the system is boosted, the wave function becomes
\begin{equation}\label{11}
\psi ^{n}_{\eta }(z,t) = \left({1 \over \pi n!2^{n}} \right)^{1/2}
H_{n} \left((e^{-\eta }u + e^{\eta }v)/\sqrt{2} \right)
\times \exp \left\{-{1\over 2}\left(e^{-2\eta }u^{2} +
e^{2\eta }v^{2}\right)\right\} .
\end{equation}


\begin{figure}[thb]  
\centerline{\psfig{figure=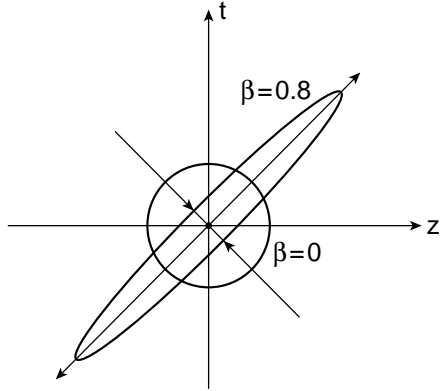,angle=0,height=60mm}}
\caption{Effect of the Lorentz boost on the space-time wave function.
The circular space-time distribution at the rest frame becomes
Lorentz-squeezed to become an elliptic distribution.}
\end{figure}


In both Eqs. (\ref{10}) and (\ref{11}), the localization property of the wave
function in the $u v$ plane is determined by the Gaussian factor, and it
is sufficient to study the ground state only for the essential feature of
the boundary condition.  The wave functions in Eq.(\ref{10}) and
Eq.(\ref{11}) then respectively become
\begin{equation}\label{13}
\psi _{0}(z,t) = \left({1 \over \pi} \right)^{1/2}
\exp \left\{-{1\over 2} (u^{2} + v^{2}) \right\} .
\end{equation}
If the system is boosted, the wave function becomes
\begin{equation}\label{14}
\psi _{\eta }(z,t) = \left({1 \over \pi }\right)^{1/2}
\exp \left\{-{1\over 2}\left(e^{-2\eta }u^{2} +
e^{2\eta }v^{2}\right)\right\} .
\end{equation}
We note here that the transition from Eq.(\ref{13}) to Eq.(\ref{14}) is a
squeeze transformation.  The wave function of Eq.(\ref{13}) is distributed
within a circular region in the $u v$ plane, and thus in the $z t$ plane.
On the other hand, the wave function of Eq.(\ref{14}) is distributed in an
elliptic region.  This ellipse is a ``squeezed'' circle with the same area
as the circle, as is illustrated in Fig.~1.

\section{Feynman's Parton Picture}\label{parton}

It is safe to believe that hadrons are quantum bound states of quarks having
localized probability distribution.  As in all bound-state cases, this
localization condition is responsible for the existence of discrete mass
spectra.  The most convincing evidence for this bound-state picture is the
hadronic mass spectra which are observed in high-energy
laboratories~\cite{fkr71,knp86}.
However, this picture of bound states is applicable only to observers in the
Lorentz frame in which the hadron is at rest.  How would the hadrons appear
to observers in other Lorentz frames?  More specifically, can we use the
picture of Lorentz-squeezed hadrons discussed in Sec.~\ref{covham}.

Proton's radius is $10^{-5}$ of that of the hydrogen atom. Therefore,
it is not unnatural to assume that the proton has a point charge in atomic
physics.  However, while carrying out experiments on electron scattering
from proton targets, Hofstadter in 1955 observed that the proton charge is
spread out.  In this experiment, an electron emits a virtual photon, which
then interacts with the proton.  If the proton consists of quarks
distributed within a finite space-time region, the virtual photon will
interact with quarks which carry fractional charges.  The scattering
amplitude will depend on the way in which quarks are distributed within the
proton.  The portion of the scattering amplitude which describes the
interaction between the virtual photon and the proton is called the form
factor.

Although there have been many attempts to explain this phenomenon within the
framework of quantum field theory, it is quite natural to expect that the
wave function in the quark model will describe the charge distribution.  In
high-energy experiments, we are dealing with the situation in which the
momentum transfer in the scattering process is large.  Indeed, the
Lorentz-squeezed wave functions lead to the correct behavior of the hadronic
form factor for large values of the momentum transfer~\cite{fuji70}.

While the form factor is the quantity which can be extracted from the
elastic scattering, it is important to realize that in high-energy
processes, many particles are produced in the final state.  They are called
inelastic processes.  While the elastic process is described by the total
energy and momentum transfer in the center-of-mass coordinate system, there
is, in addition, the energy transfer in inelastic scattering.  Therefore, we
would expect that the scattering cross section would depend on the energy,
momentum transfer, and energy transfer.  However, one prominent feature in
inelastic scattering is that the cross section remains nearly constant for a
fixed value of the momentum-transfer/energy-transfer ratio.  This phenomenon
is called ``scaling''~\cite{bj69}.

\begin{figure}[thb]  
\centerline{\psfig{figure=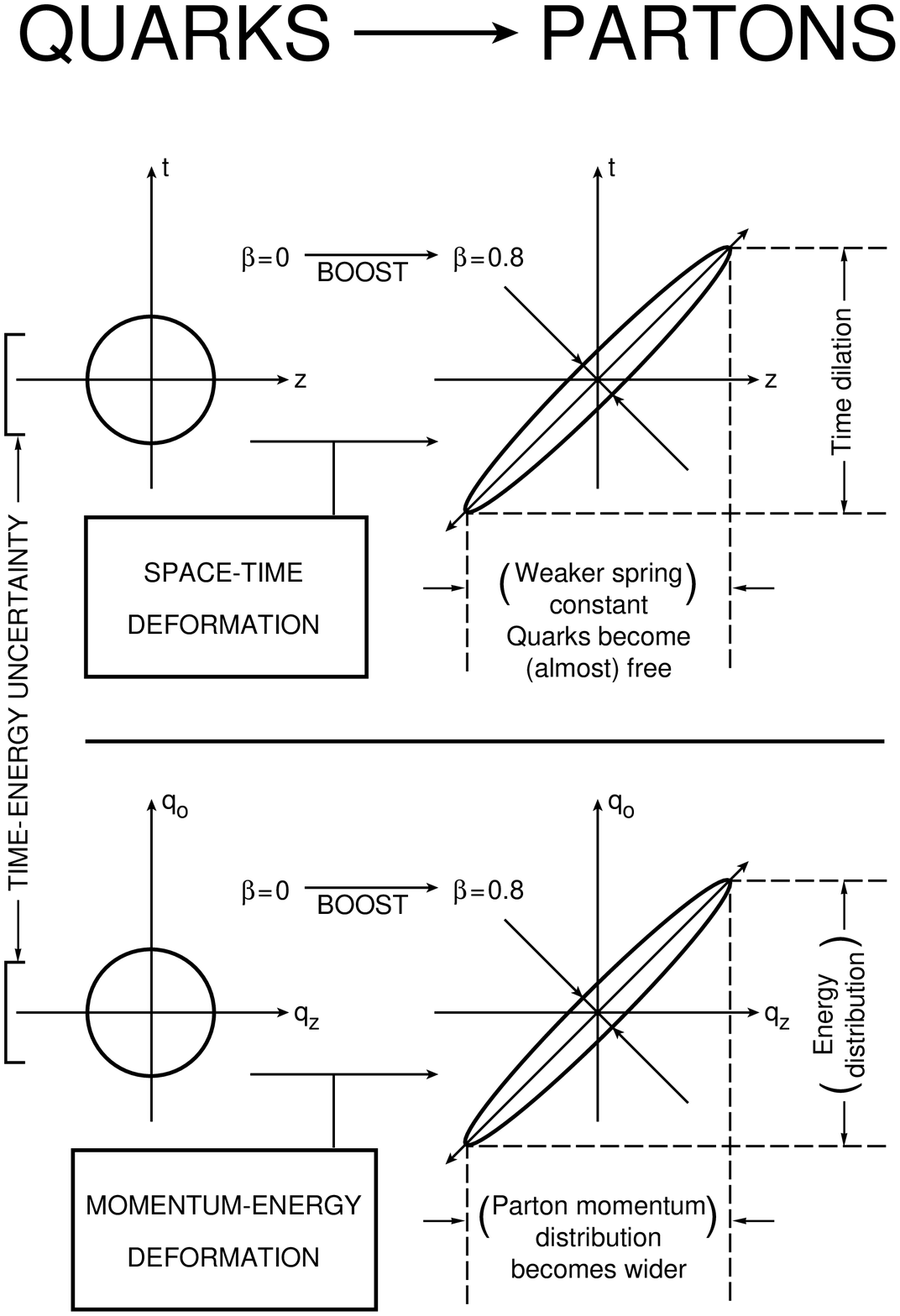,angle=0,height=140mm}}
\vspace{5mm}
\caption{Lorentz-squeezed space-time and momentum-energy wave functions.
As the hadron's speed approaches that of light, both wave functions
become concentrated along their respective positive light-cone axes.
These light-cone concentrations lead to Feynman's parton picture.}
\end{figure}

In order to explain the scaling behavior in inelastic scattering, Feynman in
1969 observed that a fast-moving hadron can be regarded as a collection of
many ``partons'' whose properties do not appear to be identical to those of
quarks~\cite{fey69}.  For example, the number of quarks inside a static
proton is three, while the number of partons in a rapidly moving proton
appears to be infinite.  The question then is how the proton looking like a
bound state of quarks to one observer can appear different to an observer in
a different Lorentz frame?  Feynman made the following systematic
observations.

    a). The picture is valid only for hadrons moving with velocity close
       to that of light.

    b). The interaction time between the quarks becomes dilated, and
        partons\\
\hspace{20mm} behave as free independent particles.

    c). The momentum distribution of partons becomes widespread as the
       hadron\\
\hspace{20mm} moves fast.

    d). The number of partons seems to be infinite or much larger than that
       of quarks.

\noindent Because the hadron is believed to be a bound state of two or three
quarks, each of the above phenomena appears as a paradox, particularly b) and
c) together.  We would like to resolve this paradox using the covariant
harmonic oscillator formalism.

For this purpose, we need a momentum-energy wave function.  If the quarks
have the four-momenta $p_{a}$ and $p_{b}$, we can construct two independent
four-momentum variables~\cite{fkr71}
\begin{equation}
P = p_{a} + p_{b} , \qquad q = \sqrt{2}(p_{a} - p_{b}) .
\end{equation}
The four-momentum $P$ is the total four-momentum and is thus the hadronic
four-momentum.  $q$ measures the four-momentum separation between the quarks.

We expect to get the momentum-energy wave function by taking the Fourier
transformation of Eq.(\ref{14}):
\begin{equation}\label{fourier}
\phi_{\eta }(q_{z},q_{0}) = \left({1 \over 2\pi }\right)
\int \psi_{\eta}(z, t) \exp{\left\{-i(q_{z}z - q_{0}t)\right\}} dx dt .
\end{equation}
Let us now define the momentum-energy variables in the light-cone coordinate
system as
\begin{equation}\label{conju}
q_{u} = (q_{0} - q_{z})/\sqrt{2} ,  \qquad
q_{v} = (q_{0} + q_{z})/\sqrt{2} .
\end{equation}
In terms of these variables, the Fourier transformation of
Eq.(\ref{fourier}) can be written as
\begin{equation}\label{fourier2}
\phi_{\eta }(q_{z},q_{0}) = \left({1 \over 2\pi }\right)
\int \psi_{\eta}(z, t) \exp{\left\{-i(q_{u} u + q_{v} v)\right\}} du dv .
\end{equation}
The resulting momentum-energy wave function is
\begin{equation}\label{phi}
\phi_{\eta }(q_{z},q_{0}) = \left({1 \over \pi }\right)^{1/2}
\exp\left\{-{1\over 2}\left(e^{-2\eta}q_{u}^{2} +
e^{2\eta}q_{v}^{2}\right)\right\} .
\end{equation}
Because we are using here the harmonic oscillator, the mathematical form
of the above momentum-energy wave function is identical to that of the
space-time wave function.  The Lorentz squeeze properties of these wave
functions are also the same, as are indicated in Fig.~2.

When the hadron is at rest with $\eta = 0$, both wave functions behave like
those for the static bound state of quarks.  As $\eta$ increases, the wave
functions become continuously squeezed until they become concentrated along
their respective positive light-cone axes.  Let us look at the z-axis
projection of the space-time wave function.  Indeed, the width of the quark
distribution increases as the hadronic speed approaches that of the speed of
light.  The position of each quark appears widespread to the observer in the
laboratory frame, and the quarks appear like free particles.

Furthermore, interaction time of the quarks among themselves become dilated.
Because the wave function becomes wide-spread, the distance between one end
of the harmonic oscillator well and the other end increases as is indicated
in Fig.~2.  This effect, first noted by Feynman~\cite{fey69}, is universally
observed in high-energy hadronic experiments.  The period is oscillation is
increases like $e^{\eta}$.  On the other hand, the interaction time with
the external signal, since it is moving in the direction opposite to the
direction of the hadron, it travels along the negative light-cone axis.  If
the hadron contracts along the negative light-cone axis, the interaction time
decreases by $e^{-\eta}$.  The ratio of the interaction time to the
oscillator period becomes $e^{-2\eta}$.  The energy of each proton coming
out of the Fermilab accelerator is $900 GeV$.  This leads the ratio to
$10^{-6}$.  This is indeed a small number.  The external signal is not able
to sense the interaction of the quarks among themselves inside the hadron.

The momentum-energy wave function is just like the space-time wave function.
The longitudinal momentum distribution becomes wide-spread as the hadronic
speed approaches the velocity of light.  This is in contradiction with our
expectation from nonrelativistic quantum mechanics that the width of the
momentum distribution is inversely proportional to that of the position wave
function.  Our expectation is that if the quarks are free, they must have
their sharply defined momenta, not a wide-spread distribution.  This apparent
contradiction presents to us the following two fundamental questions:

   a).  If both the spatial and momentum distributions become widespread
      as the hadron moves, and if we insist on Heisenberg's uncertainty
      relation, is Planck's constant dependent on the hadronic velocity?

   b).  Is this apparent contradiction related to another apparent
      contradiction that the number of partons is infinite while there
      are only two or three quarks inside the hadron?

The answer to the first question is ``No'', and that for the second question
is ``Yes''.  Let us answer the first question which is related to the Lorentz
invariance of Planck's constant.  If we take the product of the width of the
longitudinal momentum distribution and that of the spatial distribution, we
end up with the relation
\begin{equation}
<z^{2}><q_{z}^{2}> = (1/4)[\cosh(2\eta)]^{2}  .
\end{equation}
The right-hand side increases as the velocity parameter increases.  This
could lead us to an erroneous conclusion that Planck's constant becomes
dependent on velocity.  This is not correct, because the longitudinal
momentum variable $q_{z}$ is no longer conjugate to the longitudinal
position variable when the hadron moves.

In order to maintain the Lorentz-invariance of the uncertainty product,
we have to work with a conjugate pair of variables whose product does
not depend on the velocity parameter.  Let us go back to Eq.(\ref{conju})
and Eq.(\ref{fourier2}).  It is quite clear that the light-cone variable
$u$ and $v$ are conjugate to $q_{u}$ and $q_{v}$ respectively.  It is
also clear that the distribution along the $q_{u}$ axis shrinks as the
$u$-axis distribution expands.  The exact calculation leads to
\begin{equation}
<u^{2}><q_{u}^{2}> = 1/4 , \qquad  <v^{2}><q_{v}^{2}> = 1/4  .
\end{equation}
Planck's constant is indeed Lorentz-invariant.

Let us next resolve the puzzle of why the number of partons appears to
be infinite while there are only a finite number of quarks inside the
hadron.  As the hadronic speed approaches the speed of light, both the
x and q distributions become concentrated along the positive light-cone
axis.  This means that the quarks also move with velocity very close
to that of light.  Quarks in this case behave like massless particles.

We then know from statistical mechanics that the number of massless
particles is not a conserved quantity.  For instance, in black-body
radiation, free light-like particles have a widespread momentum
distribution.  However, this does not contradict the known principles
of quantum mechanics, because the massless photons can be divided into
infinitely many massless particles with a continuous momentum
distribution.

Likewise, in the parton picture, massless free quarks have a wide-spread
momentum distribution.  They can appear as a distribution of an
infinite number of free particles.  These free massless particles are the
partons.  It is possible to measure this distribution in high-energy
laboratories, and it is also possible to calculate it using the covariant
harmonic oscillator formalism.  We are thus forced to compare these two
results.  Indeed, according to Hussar's calculation~\cite{hussar81},
the Lorentz-boosted oscillator wave function produces a reasoanbly
accurate parton distribution.

\section*{Concluding Remarks}
The phenomenological aspects of the covariant oscillator formalism have
been extensively discussed in the literature~\cite{knp86}.  The purpose
of the present paper was to put the formalism into its proper place
in the Lorentz-covariant world of physics.

We have shown that the oscillator formalism constitutes a
representation of Wigner's little group governing the internal space-time
symmetries of relativistic particles.  For this purpose, we have given
a comprehensive review of the little groups for massive and massless
particles.  We have discussed also the contraction procedure in which
the $E(2)$-like little group for massless particles is obtained from
the $O(3)$-like little group for massive particles.  We have given a
comprehensive review of the contents of Table I.

\section*{Acknowledgments}
Since I came to the United States in 1954 right after high-school
graduation, I made many new friends from many different countries.  I
have benefitted greatly from my association with Chinese friends and
colleagues whose backgrounds are very similar to mine.  When I
met Professor C. N. Yang in 1958, he told me about the period before
1945 when he and other Chinese students had to move to a south-western
province of China to continue their studies.  Professor Yang's story
was a great encouragement to me since I had a similar experience
during the Korean conflict which lasted from 1950 to 1953.  In
Appendix \ref{kant}, I discuss how my Eastern background influenced
the research program which I outlined in this paper.

I am indeed honored to join my Chinese colleagues in celebrating the
90th birth year of Professor T. Y. Wu.  I am particularly grateful
to Professor J. P Hsu for inviting me to write this review article on
the covariant harmonic oscillators.

\begin{appendix}

\section{Contraction of O(3) to E(2)}\label{o3e2}
In this Appendix, we explain what the $E(2)$ group is.  We then
explain how we can obtain this group from the three-dimensional
rotation group by making a flat-surface or cylindrical approximation.
This contraction procedure will give a clue to obtaining the $E(2)$-like
symmetry for massless particles from the $O(3)$-like symmetry for
massive particles by making the infinite-momentum limit.

The $E(2)$ transformations consist of rotation and two translations on
a flat plane.  Let us start with the  rotation matrix applicable to
the column vector $(x, y, 1)$:
\begin{equation}\label{rot}
R(\theta) = \pmatrix{\cos\theta & -\sin\theta & 0 \cr
\sin\theta & \cos\theta & 0 \cr 0 & 0 & 1} .
\end{equation}
Let us then consider the translation matrix:
\begin{equation}
T(a, b) = \pmatrix{1 & 0 & a \cr 0 & 1 & b \cr 0 & 0 & 1} .
\end{equation}
If we take the product $T(a, b) R(\theta)$,
\begin{equation}\label{eucl}
E(a, b, \theta) = T(a, b) R(\theta) =
\pmatrix{\cos\theta & -\sin\theta & a \cr
\sin\theta & \cos\theta & b \cr 0 & 0 & 1} .
\end{equation}
This is the Euclidean transformation matrix applicable to the
two-dimensional $x y$ plane.  The matrices $R(\theta)$ and $T(a,b)$
represent the rotation and translation subgroups respectively.  The
above expression is not a direct product because $R(\theta)$ does not
commute with $T(a,b)$.  The translations constitute an Abelian invariant
subgroup because two different $T$ matrices commute with each other,
and because
\begin{equation}
R(\theta) T(a,b) R^{-1}(\theta) = T(a',b') .
\end{equation}
The rotation subgroup is not invariant because the conjugation
$$T(a,b) R(\theta) T^{-1}(a,b)$$
does not lead to another rotation.

We can write the above transformation matrix in terms of generators.
The rotation is generated by
\begin{equation}
J_{3} = \pmatrix{0 & -i & 0 \cr i & 0 & 0 \cr 0 & 0 & 0} .
\end{equation}
The translations are generated by
\begin{equation}
P_{1} = \pmatrix{0 & 0 & i \cr 0 & 0 & 0 \cr 0 & 0 & 0} , \qquad
P_{2} = \pmatrix{0 & 0 & 0 \cr 0 & 0 & i \cr 0 & 0 & 0} .
\end{equation}
These generators satisfy the commutation relations:
\begin{equation}\label{e2com}
[P_{1}, P_{2}] = 0 , \qquad [J_{3}, P_{1}] = iP_{2} ,
\qquad [J_{3}, P_{2}] = -iP_{1} .
\end{equation}
This $E(2)$ group is not only convenient for illustrating the groups
containing an Abelian invariant subgroup, but also occupies an
important place in constructing representations for the little
group for massless particles, since the little group for massless
particles is locally isomorphic to the above $E(2)$ group.

The contraction of $O(3)$ to $E(2)$ is well known and is often called
the Inonu-Wigner contraction~\cite{inonu53}.  The question is whether
the $E(2)$-like little group can be obtained from the $O(3)$-like
little group.  In order to answer this question, let us closely look
at the original form of the Inonu-Wigner contraction.  We start with
the generators of $O(3)$.  The $J_{3}$ matrix is given in Eq.(\ref{j3}),
and
\begin{equation}\label{o3gen}
J_{2} = \pmatrix{0&0&i\cr0&0&0\cr-i&0&0} , \qquad
J_{3} = \pmatrix{0&-i&0\cr i &0&0\cr0&0&0} .
\end{equation}
The Euclidean group $E(2)$ is generated by $J_{3}, P_{1}$ and $P_{2}$,
and their Lie algebra has been discussed in Sec.~\ref{intro}.

Let us transpose the Lie algebra of the $E(2)$ group.  Then $P_{1}$ and
$P_{2}$ become $Q_{1}$ and $Q_{2}$ respectively, where
\begin{equation}
Q_{1} = \pmatrix{0&0&0\cr0&0&0\cr i &0&0} , \qquad
Q_{2} = \pmatrix{0&0&0\cr0&0&0\cr0&i&0} .
\end{equation}
Together with $J_{3}$, these generators satisfy the
same set of commutation relations as that for
$J_{3}, P_{1}$, and $P_{2}$ given in Eq.(\ref{e2com}):
\begin{equation}
[Q_{1}, Q_{2}] = 0 , \qquad [J_{3}, Q_{1}] = iQ_{2} , \qquad
[J_{3}, Q_{2}] = -iQ_{1} .
\end{equation}
These matrices generate transformations of a point on a circular
cylinder.  Rotations around the cylindrical axis are generated by
$J_{3}$.  The matrices $Q_{1}$ and $Q_{2}$ generate translations along
the direction of $z$ axis.  The group generated by these three matrices
is called the {\it cylindrical group}~\cite{kiwi87jm,kiwi90jm}.

We can achieve the contractions to the Euclidean and cylindrical groups
by taking the large-radius limits of
\begin{equation}\label{inonucont}
P_{1} = {1\over R} B^{-1} J_{2} B ,
\qquad P_{2} = -{1\over R} B^{-1} J_{1} B ,
\end{equation}
and
\begin{equation}
Q_{1} = -{1\over R}B J_{2}B^{-1} , \qquad
Q_{2} = {1\over R} B J_{1} B^{-1} ,
\end{equation}
where
\begin{equation}\label{bmatrix}
B(R) = \pmatrix{1&0&0\cr0&1&0\cr0&0&R}  .
\end{equation}
The vector spaces to which the above generators are applicable are
$(x, y, z/R)$ and $(x, y, Rz)$ for the Euclidean and cylindrical groups
respectively.  They can be regarded as the north-pole and equatorial-belt
approximations of the spherical surface respectively~\cite{kiwi87jm}.

\section{Contraction of O(3)-like Little Group to E(2)-like Little
Group}\label{contrac}

Since $P_{1} (P_{2})$ commutes with $Q_{2} (Q_{1})$, we can consider the
following combination of generators.
\begin{equation}
F_{1} = P_{1} + Q_{1} , \qquad F_{2} = P_{2} + Q_{2} .
\end{equation}
Then these operators also satisfy the commutation relations:
\begin{equation}\label{commuf}
[F_{1}, F_{2}] = 0 , \qquad [J_{3}, F_{1}] = iF_{2} , \qquad
[J_{3}, F_{2}] = -iF_{1} .
\end{equation}
However, we cannot make this addition using the three-by-three matrices
for $P_{i}$ and $Q_{i}$ to construct three-by-three matrices for $F_{1}$
and $F_{2}$, because the vector spaces are different for the $P_{i}$ and
$Q_{i}$ representations.  We can accommodate this difference by creating
two different $z$ coordinates, one with a contracted $z$ and the other
with an expanded $z$, namely $(x, y, Rz, z/R)$.  Then the generators
become
\begin{equation}
P_{1} = \pmatrix{0&0&0&i\cr0&0&0&0\cr0&0&0&0\cr0&0&0&0} , \qquad
P_{2} = \pmatrix{0&0&0&0\cr0&0&0&i\cr0&0&0&0\cr0&0&0&0} .
\end{equation}
\begin{equation}
Q_{1} = \pmatrix{0&0&0&0\cr0&0&0&0\cr i &0&0&0\cr0&0&0&0} , \qquad
Q_{2} = \pmatrix{0&0&0&0\cr0&0&0&0\cr0&i&0&0\cr0&0&0&0} .
\end{equation}
Then $F_{1}$ and $F_{2}$ will take the form
\begin{equation}\label{f1f2}
F_{1} = \pmatrix{0&0&0&i\cr0&0&0&0\cr i &0&0&0\cr0&0&0&0} , \qquad
F_{2} = \pmatrix{0&0&0&0\cr0&0&0&i\cr0&i&0&0\cr0&0&0&0} .
\end{equation}
The rotation generator $J_{3}$ takes the form of Eq.(\ref{j3}).
These four-by-four matrices satisfy the E(2)-like commutation relations
of Eq.(\ref{commuf}).

Now the $B$ matrix of Eq.(\ref{bmatrix}), can be expanded to
\begin{equation}\label{bmatrix2}
B(R) = \pmatrix{1&0&0&0\cr0&1&0&0\cr0&0&R&0\cr0&0&0&1/R} .
\end{equation}
If we make a similarity transformation on the above form using the matrix
\begin{equation}\label{simil}
\pmatrix{1&0&0&0\cr0&1&0&0\cr0&0&1/\sqrt{2} &-1/\sqrt{2}
\cr0&0&1/\sqrt{2}&1/\sqrt{2}} ,
\end{equation}
which performs a 45-degree rotation of the third and fourth coordinates,
then this matrix becomes
\begin{equation}\label{simil2}
\pmatrix{1&0&0&0\cr0&1&0&0\cr0&0 & \cosh\eta & \sinh\eta
\cr0 & 0 & \sinh\eta & \cosh\eta} ,
\end{equation}
with $R = e^\eta$.  This form is the Lorentz boost matrix along the $z$
direction.  If we start with the set of expanded rotation generators
$J_{3}$ of Eq.(\ref{j3}), and
perform the same operation as the original Inonu-Wigner contraction
given in Eq.(\ref{inonucont}), the result is
\begin{equation}
N_{1} = {1\over R} B^{-1} J_{2} B ,
\qquad N_{2} = -{1\over R} B^{-1} J_{1} B ,
\end{equation}
where $N_{1}$ and $N_{2}$ are given in Eq.(\ref{n1n2}).  The generators
$N_{1}$ and $N_{2}$ are the contracted $J_{2}$ and $J_{1}$ respectively
in the infinite-momentum/zero-mass limit.

\section{Covariance and Its Historical Background}\label{kant}
Unlike classical physics, modern physics depends heavily on observer's
state of mind or environment.  In special relativity, observers in
different Lorentz frames see the
same physical system differently.  The importance of the observer's
subjective viewpoint was emphasized by Immanuel Kant in his book
entitled {\it Kritik der reinen Vernunft} whose first and second
editions were published in 1781 and 1787 respectively.  However, using
his own logic, he ended up with a conclusion that there must be the
absolute inertial frame, and that we only see the frames dictated by
our subjectivity.

Einstein's special relativity was developed along Kant's line of thinking:
things depend on the frame from which you make observations.  However,
there is one big difference.  Instead of the absolute frame, Einstein
introduced an extra dimension.  Let us illustrate this using a CocaCola
can.  It appears like a circle if you look at it from the top, while
it appears as a rectangle from the side.  The real thing is a
three-dimensional circular cylinder.  While Kant was obsessed with
the absoluteness of the real thing, Einstein was able to observe the
importance of the extra dimension.

I was fortunate enough to be close to Eugene Wigner, and enjoyed the
privilege of asking him many questions.  I once asked him whether he
thinks like Immanuel Kant.  He said Yes.  I then asked him whether
Einstein was a Kantianist in his opinion.  Wigner said very firmly Yes.
I then asked him whether he studied the philosophy of Kant while he
was in college.  He said No, and said that he realized he had been a
Kantianist after writing so many papers in physics.  He added that
philosophers do not dictate people how to think, but their job is
to describe systematically how people think.  Wigner told me that I
was the only one who asked him this question, and asked me how I knew
the Kantian way of reasoning was working in his mind.  I gave him the
following answer.

I never had any formal education in oriental philosophy, but I know
that my frame of thinking is affected by my Korean background.  One
important aspect is that Immanuel Kant's name is known to every
high-school graduate in Korea, while he is unknown to Americans,
particularly to American physicists.  The question then is whether
there is in Eastern culture a ``natural frequency'' which can
resonate with one of the frequencies radiated from Kantianism developed
in Europe.

I would like to answer this question in the following way.  Koreans
absorbed a bulk of Chinese culture during the period of the Tang
dynasty (618-907 AD).  At that time, China was the center of the world
as the United States is today.  This dynasty's intellectual life was
based on Taoism which tells us, among others, that everything in this
universe has to be balanced between its plus (or bright) side and its
minus (or dark) side.  This way of thinking forces us to look at things
from two different or opposite directions.  This aspect of Taoism could
constitute a ``natural frequency'' which can be tuned to the Kantian
view of the world where things depend how they are observed.

I would like to point out that Hideki Yukawa was quite fond of Taoism and
studied systematically the books of Laotse and Chuangtse who were the
founding fathers of Taoism~\cite{tani79}.  Both Laotse and Chuangtse
lived before the time of Confucius.  It is interesting to note that
Kantianism is also popular is Japan, and it is my assumption that Kant's
books were translated into Japanese by Japanese philosophers first,
and Koreans of my father's age learned about Kant by reading the
translated versions.
My publication record will indicate that I studied Yukawa's papers
before becoming seriously interested in Wignerism.  Indeed, I picked up
a signal of possible connection between Kantianism and Taoism while
reading Yukawa's papers carefully, and this led to my bold venture to
ask Wigner whether he was a Kantianist.

Kant wrote his books in German, but he was born and spent his entire
life in a Baltic enclave now called Kaliningrad located between Poland
and Lithuania.  Historically, this place was dominated by several
different countries with different ideologies~\cite{apple94}.  However,
Kant's view was that the people there may appear differently depending
on who look at them, but they remain unchanged.  At the same time,
they had to entertain different ideologies imposed by different rulers.
Kant translated this philosophy into physics when he discussed the
absolute and relative frames.  He was obsessed with the absolute frame,
and this is the reason why Kant is not regarded as a physicist in
Einstein's world in which we live.

The people of Kant's land stayed in the same place while experiencing
different ideological environments.  Almost like Kant, I was exposed
to two different cultural environments by moving myself from Asia to
the United States.  Thus, I often had to raise the question of
absolute and relative values.  Let us discuss this problem using one
concrete example.

About 4,500 years ago, there was a king named Yao in China.  While he
was looking for a man who could serve as the prime minister, he heard
from many people that a person named Shiyu was widely respected and
had a deep knowledge of the world.  The king then sent his messengers
to invite Shiyu to come to his palace and to run the country.  After
hearing the king's message, Shiyu without saying anything went to a
creek in front of his house and started washing his ears.  He thought
he heard the dirtiest story in his life.

Shiyu is still respected in the Eastern world as one of the wisest men
in history.  We do not know whether this person existed or is a made-up
personality.  In either case, we are led to look for a similar person
in the Western world.  In ancient Greece, each city was run by its
city council.  As we
experience even these days, people accomplish very little in committee
meetings.  Thus, it is safe to assume that the city councils in ancient
Greece did not handle matters too efficiently.  For this reason, there
was a well-respected wiseman like Shiyu who never attended his city
council meetings.  His name was Idiot.  Idiot was a wiseman, but he
never contributed his wisdom to his community.  His fellow citizens
labeled him as a useless person.  This was how the word idiot was
developed in the Western world.

Idiot and Shiyu had the same personality if they were not the same
person.  However, Idiot is a useless person in state-centered societies
like Sparta.  The same person is regarded as the ultimate wiseman in a
self-centered society like Korea.  I cannot say that I know everything
about other Asian countries, but I have a deep knowledge of Korea where
I was born and raised.  The same person looks quite differently to
observers in different cultural frames.  While doing research in the
United States with my Eastern background, I was frequently forced to
find a common ground for two seemingly different views.  This cultural
background strongly influenced me in producing the further contents of
Einstein's $E = mc^{2}$ tabulated in Table I~\cite{kim89}.

Let us go back to the question of relative values.  For Taoists, those
two opposite faces of the same person is like ``yang'' (plus) versus
``ying'' (minus).  Finding the harmony between these two opposite
points of view is the ideal way to live in this world.  We cannot
always live like Shiyu, nor like Idiot.  The key to happiness is to
find a harmony between the individual and the society to which
he/she belongs.  The key word here seems to be ``harmony.''

To Kantianists, however, it is quite natural for the same character
to appear differently in two different environments.  The problem is
to find the absolute value from these two different faces.  Does this
absolute value exist?  According to Kant, it exists.  To most of us,
it is very difficult to find it if it exists.

Let us finally visit Einstein.  He avoids the question of the existence
of the absolute value.  Instead, he introduces a new variable.  The
variable is the ratio between the individual's ability to contribute
and the community's need for his service.  The best way to live in this
world is to adjust this variable to the optimal value.  Einstein's
approach is to a quantification of Taoism by introducing a new
variable.

If Taoism is so close to Einsteinism, why do we have to mention Kant
at all?  We have to keep in mind that Kant was the first person who
formulated the idea that observers can participate in drawing the
picture of the world.  It is not clear whether Einstein could have
formulated his relativity theory without Kant.  Indeed, Kant spent many
years for studying physics, namely observer-dependent physics.  However,
because of his obsession toward the absolute thing, he spent all of
his time for finding the absolute frame.  If one has a Taoist background,
he/she is more likely to appreciate the concept of relativistic
covariance.

I would like to stress that Taoism is not confined to the ancient
Eastern world.  It is practiced frequently in the United States.
Let us look at American football games.  The offensive strategy does
not rely on brute force, but is aimed at breaking the harmony of the
defense.  For instance, when the offensive team is near the end zone,
the defense becomes very strong because it covers only a small area.
Then, it is not uncommon for the offense to place four wide-receivers
instead of two.  This will divide the defense into two sides while
creating a hole in the middle.  Then the quarter-back can carry the
ball to the end zone.  The key word is to destroy the balance of the
defense.

Taoism forms the philosophical base for Sun Tzu's classic book on
military arts~\cite{suntzu96}.  When I watch the football games, I
watch them as Sun-Tzu games.  My maternal grandfather was fluent in
the Chinese classic literature, and he was particularly fond of Sun Tzu.
He told me many stories from Sun Tzu's books.  This presumably was
how I inherited some of the Taoist tradition.  As I said in this paper,
my research life was influenced by my Asian background.  Many of my
Asian friends complain that they are handicapped to do original
research because of the East-West cultural difference.  I disagree
with them.  This difference could be the richest source of originality.

If the concept of cultural difference is too abstract to grasp, we can
illustrate it in terms of a secular example.  Russia under the
communist rule had a uniform price system.  The collapse of communism
caused price differences for different regions.  This caused economic
hardship for many Russians.  But some creative Russian merchants
are able to accumulate fortunes by buying things from one region
and selling them in another region.  Can we blame those merchants?
Likewise, there is an enormous cultural gap between the East and the
West.  It is up to us how to make profit from this difference.

\end{appendix}

\end{document}